\documentclass[twocolumn,showpacs,amsmath,amssymb,prb]{revtex4}

\usepackage{graphicx}
\usepackage[usenames]{color}

\newcommand{\bLR}{\ensuremath f_L(\epsilon)[1-f_R(\epsilon)]}
\newcommand{\bRL}{\ensuremath [1-f_L(\epsilon)]f_R(\epsilon)}
\newcommand{\bLRs}{\ensuremath f_L(1-f_R)}
\newcommand{\bRLs}{\ensuremath (1-f_L)f_R}
\newcommand{\GL}{\ensuremath\Gamma_L}
\newcommand{\GR}{\ensuremath\Gamma_R}
\newcommand{\Dinf}{\ensuremath\mathcal{D}_{}}
\newcommand{\up}{\uparrow}
\newcommand{\dn}{\downarrow}

\begin{document}

\title{Coulomb-interaction effects in full counting statistics of a quantum-dot ABI}
\author{Daniel Urban}
\affiliation{Theoretische Physik, Universit\"at Duisburg-Essen, 47048 Duisburg\\
Institut f\"ur Theoretische Physik III, Ruhr-Universit\"at Bochum, 44780 Bochum, Germany}

\author{J\"urgen K\"onig}
\affiliation{Theoretische Physik, Universit\"at Duisburg-Essen, 47048 Duisburg\\
Institut f\"ur Theoretische Physik III, Ruhr-Universit\"at Bochum, 44780 Bochum, Germany}

\author{Rosario Fazio}
\affiliation{International School for Advanced Studies (SISSA), 34014 Trieste, Italy}
\affiliation{NEST-CNR-INFM \& Scuola Normale Superiore, 56126 Pisa, Italy}

\date{\today}

\begin{abstract}
We study the effect of Coulomb interaction on the full counting statistics of an ABI with a single-level quantum dot in one arm in the regime of weak dot-lead and lead-lead tunnel coupling.
In the absence of Coulomb interaction the interference processes are of non-resonant nature with an even AB flux dependence and obey bidirectional Poissonian statistics.
For large charging energy the statistic of these processes changes.
In addition processes of resonant nature with an odd flux dependence appear.
In the limit of strongly asymmetric tunnel couplings from the dot to the left and right leads, their statistics is found to be strongly super-Poissonian.
\end{abstract}

\maketitle

\section{Introduction}

The transport through mesoscopic systems is fully described by the full counting statistics (FCS), i.e.~knowledge of the current distribution function.\cite{levitov:1996}
Perseverant theoretical efforts have brought about several formalisms for the determination of the cumulant generating function (CGF) also in the presence of interaction\cite{QNmesPhys,nazarov:2002,boltzmannlangevin,stochasticPathint,KeldyshGF,braggio:2006,flindt:2008}
and with an extension to frequency-dependent FCS.\cite{emary:2007,flindt:2008} A large variety of systems containing quantum dots (QDs) have been treated considering the tunneling limit,\cite{TunnelDots,bagrets:2003,kiesslich:2006} the Kondo regime,\cite{gogolin:2006,schmidt:2007} superconducting leads\cite{SCleads} as well as completely superconducting systems\cite{choi:2001,heikkila}, and nano-mechanical resonators.\cite{NEMS}

Interest in FCS is founded on the fact that it offers more information than average current and noise: The Fano factor is known to relate to the average number of transferred charges\cite{fanogreaterone} (as measured for instance in Andreev reflection\cite{andreevreflection} or Quantum Hall Effect\cite{fanolessone}). However, in case the current is caused by several elementary events with different numbers of transferred charges, it does not uniquely identify these. In this case it is helpful to identify the individual processes directly in the cumulant generating function.\cite{cuevas:2003,johansson:2003,belzig:2005,vanevic:2007}

The measurement of higher moments of the current correlator with spectrum analyzers requires long averaging times and and may cause significant environmental backaction.\cite{hist3rdmoment}
The extension of methods for the detection of noise employing qubits\cite{qubitdetector} and Josephson junctions\cite{thermalescape,cooperpairtunneling} also allows the detection of lower-order moments (up to the fourth\cite{MQTfourthcumulant}).
A more direct measurement of the FCS is possible with Josephson junctions as threshold detectors.\cite{thresholddetectors}
The most direct technique currently available relies on detection of individual charges:
Electrons traversing a QD can be detected by means of a nearby Quantum Point Contact (QPC).\cite{gustavsson:2006,fricke:2007} For a single dot this approach is limited to the shot-noise regime. Recently this restriction was overcome in measuring the time trace of the point-contact current in the vicinity of two quantum dots in series.\cite{fujisawa:2006}
With this technique it is possible to directly measure the probability distribution $P(N,t_0)$ that $N$ charges have passed through the system after a given time $t_0$.
All cumulants of the current can be obtained from the CGF
\begin{equation}
  S (\chi) = \ln \left[ \sum_{N=-\infty}^\infty e^{iN\chi}P(N,t_0) \right]
\end{equation}
by performing derivatives with respect to the counting field
$\kappa^{(n)}=(-i)^n( e^n/t_0) \partial^n_\chi
S(\chi)|_{\chi=0}$.

In this work we want to address yet a different problem related to FCS--its properties in
multiply connected geometries in the presence of strong local correlations. AB (AB)
interferometry of electronic transport through multiply-connected mesoscopic devices offers the possibility to probe coherence of transport channels via interference of different paths enclosing a magnetic flux.
One important issue in the context of AB interferometry is how interaction introduces dephasing.
For the special case of Mach-Zehnder interferometers, different choices of dephasing probes have been analyzed in the context of FCS.\cite{forster}
In interferometers with embedded quantum dots dephasing can also be introduced by detecting the electrons on the dot\cite{buks:1998,khym:2006,ChargeDetector,EdgeStates} or their phase.\cite{PhaseDetector}
Similar effects have also been observed in the Kondo regime.\cite{KondoDephasing}
Here we study the FCS of electronic transport through an ABI with
a quantum dot embedded in one of the interferometer arms.
Such a system has been analyzed both experimentally\cite{PhaseLapseOriginal,PhaseLapseNewExps1,PhaseLapseNewExps2,kobayashi:2004,unexpectedperiodicity,buks:1998} and theoretically\cite{PhaseLapseAnalysis,hofstetter,bruder:1996,pala:2004} in a large number of publications.

The main question to be addressed in this paper is how Coulomb interaction, giving rise to a large charging energy in the QD, affects the coherence of the transport processes, probed by interference visible in oscillations of the current or the current noise as a function of the enclosed magnetic flux (AB oscillations).
Thereby we aim at identifying the underlying transport mechanisms by means of the FCS rather than just analyzing how they affect current or current noise.

\begin{figure}\center
    \includegraphics[width=.8\columnwidth]{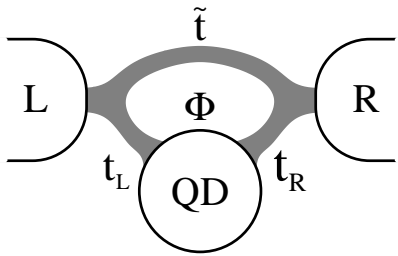}
  \caption{The quantum-dot ABI: Two equilibrium reservoirs are connected by a direct tunneling path and one in which a quantum dot is embedded. Together the paths enclose a magnetic flux $\Phi$.}
  \label{fig:system}
\end{figure}

We consider the quantum-dot ABI shown in
Fig.~\ref{fig:system}: one arm of the interferometer contains a single-level
quantum dot, the other one provides a direct tunneling path.
The two arms enclose a magnetic flux $\Phi$.
For reference, we first summarize the situation in the absence of
Coulomb interaction. In this case, a single-particle treatment of noninteracting electrons can
be used. The full counting statistics for transport of noninteracting electrons
through an arbitrary two-terminal setup is given by the Levitov-Lesovik formula,\cite{levitov:1996}
\begin{eqnarray}
  S(\chi)   &=& 2 t_0 \int \frac{d\omega}{2\pi\hbar} \ln
  \Bigl\{ 1 + T(\omega) \bigl[ f_L(\omega)[1-f_R(\omega)](e^{i\chi}-1)
\nonumber\\
    & &  + [1-f_L(\omega)]f_R(\omega)(e^{-i\chi}-1)  \bigr]\Bigr\}
\end{eqnarray}
which expresses the cumulant generating function in terms of the
energy-dependent transmission probability $T(\omega)$ per spin channel,
where $f_L(\omega)$ and $f_R(\omega)$ are the Fermi functions of the left and
right leads, respectively.
As in this paper we are going to consider a perturbative expansion in the
tunneling: let us consider only the lowest non-trivial orders for the noninteracting
case as well.

The transmission probability is the modulus squared of the transmission
amplitude that, for the quantum-dot ABI,
is the sum of the amplitudes for transmission through the reference arm and
through the arm containing the quantum dot, respectively.
Taking the exact transmission from Ref.~[\onlinecite{kubala}], plugging it into the
Levitov-Lesovik formula, and then expanding the resulting expression to lowest order in both tunnel-coupling strengths through the reference arm and the dot (corresponding to electron paths encircling the flux once) leads to the following result:
\begin{eqnarray}\label{eq:S01}
     S_0(\chi,\phi)  &\sim&		\tilde{t}\, t_L t_R \cos{\phi}	\nonumber \\
				&\times&	\left[ \int' \,\frac{\mathrm{d}\omega}{\pi} \frac{f_L(\omega)[1-f_R(\omega)]}{\epsilon-\omega} \,(e^{i\chi}-1) \right. \nonumber\\
				&    +    &	\left. \int' \,\frac{\mathrm{d}\omega}{\pi} \, \frac{[1-f_L(\omega)]f_R(\omega)}{\epsilon-\omega} \,(e^{-i\chi}-1) \right] \, ,
\end{eqnarray}
with the transmission amplitudes $\tilde{t},t_L,t_R$, the AB-flux $\phi$, level position $\epsilon$ and the prime denoting Cauchy's principal value.

The transport processes giving rise to the flux dependence are interference
contributions of direct tunneling between the leads and elastic (i.e.
energy- and spin-conserving) cotunneling through the single-level quantum dot.
They are proportional to the transmission
amplitude through the reference arm and to the transmission rate through the QD.
The main properties of the noninteracting case (to be contrasted later with our
results in the presence of electron correlations) are:\\
(i) The statistics is {\em bidirectional Poissonian}.
This indicates that the individual transport processes are independent from
each other, which is consistent with the notion that elastic cotunneling
through the quantum dot does not change its state.\\
(ii) The cumulant generating function is an {\em even} function of the
AB phase $\phi$.
Onsager relations\cite{onsager} require the linear conductance of a
two-terminal device to be even in the magnetic field.
For noninteracting electrons, the transmission for each energy, that could
be probed by the linear conductance at low temperature, is independent from
the other energies.
Therefore, in the absence of Coulomb interaction, the cumulant generating
function has to be an even function of $\phi$ in leading order of the
transmission.\\
(iii) As the amplitude for transmission through the quantum dot is determined by cotunneling, the energy $\omega$ of the incoming electron does not have to match the energy $\epsilon$ of the quantum level.
Therefore the processes are of {\em off-resonant} nature.

In the presence of Coulomb interaction, the picture changes qualitatively.
It is well known that interference of electrons may be affected strongly by
Coulomb interaction.
This is also the case for the model considered in this paper, an AB
interferometer, in which a single-level quantum dot with large charging energy
is embedded.
It has been predicted\cite{koeniggefen} and experimentally
confirmed\cite{kobayashi:2004} that the interference signal probed by the
linear conductance is partially reduced in the parameter regime where the quantum dot is predominantly singly occupied.
The reason is that those electrons transferred through the dot that flip
their spin with the spin of the quantum-dot electron do not contribute
to the interference signal, while the nonspin-flip processes do contribute.

It is therefore natural to expect that also the full counting statistics
will display this partial reduction of the interference signal due to spin-flip
processes.
Furthermore, there is no reason why any of the three properties formulated
above for noninteracting electrons should still hold in the presence of a large
charging energy.
In fact, we will find that all of them are changed.
In particular, we will demonstrate that {\em on-resonant} processes
that give rise to an {\em odd} $\phi$-dependence of the cumulant generating
function occur and that the FCS is not bidirectional Poissonian any more.
Onsager relations are not violated since these on-resonant processes do not contribute to the linear conductance (and equilibrium noise).
These processes have the interesting property that, for asymmetric tunnel
couplings of the quantum dot to the left and right lead, they display enhanced generalized Fano factors that can be associated with the transfer of double charges.

The paper is structured as follows:
In Section~\ref{sec:system} we introduce the model of our system and recapitulate how to obtain the FCS starting from a generalized master equation.
In Section~\ref{sec:nonint} we derive the cumulant generating function for a single-level quantum dot with large charging energy and discuss how it is modified as compared to the noninteracting limit.
Different processes are identified whose properties in the shot-noise regime are further analyzed in Section~\ref{sec:SN}.
Since the FCS of transport through a quantum dot can be measured by making use of a quantum point contact as a detector for the quantum-dot charge, it is interesting to know whether and to what extent the properties of the FCS for the quantum-dot ABI can be probed by attaching such a quantum-point-contact detector.
This is done in Section~\ref{sec:detector}.
Finally we conclude in Sec.~\ref{sec:conclusion}.

\section{System}    \label{sec:system}

The quantum-dot ABI shown in
Fig.~\ref{fig:system} is described by the following Hamiltonian:
\begin{equation}\label{eq:hamiltonian}
   H = H_\text{dot} + H_\text{leads} + H_{T,\text{dot}} + H_{T,\text{ref}}
\, .
\end{equation}
The quantum dot, $H_\text{dot} = \sum_\sigma \varepsilon\, c_\sigma^\dagger
c_\sigma + U n_{\uparrow}n_{\downarrow}$, is described by an Anderson impurity
with a spin-degenerate electronic level $\epsilon$ and charging energy $U$
for double occupation.
The leads are described as reservoirs of noninteracting fermions $ H_\text{leads} = \sum_{r,k,\sigma} \varepsilon_{rk\sigma}^{~}\, a_{rk\sigma}^\dagger a_{rk\sigma} $ with indices for lead $r \in \{L,R\}$, momentum $k$ and spin $\sigma\in\{\uparrow,\downarrow\}$. The tunneling Hamiltonian consists of two parts
\begin{eqnarray}
    H_{T,\text{dot}} &=&    \sum_{rk\sigma}     \,t_r\,       d^\dag_\sigma c_{rk\sigma}^{} + \text{h.c.}        \\
    H_{T,\text{ref}} &=&    \sum_{kk'\sigma} \,\tilde{t}\, c_{Rk\sigma}^\dag c^{}_{Lk'\sigma}    + \text{h.c.}        .
\end{eqnarray}
The first part describes tunneling between dot and leads, while the second accounts for the reference arm. We choose a gauge in which the AB phase $\phi=2\pi\Phi/\Phi_0$ is incorporated in the phase of the tunneling amplitude
$ \tilde{t} = |\tilde{t}| \, e^{i \phi} $.

The strength of the tunnel coupling to the dot is characterized by the transition rates $ \hbar \Gamma_r(\omega) =  2 \pi \sum_k \left|t_{rk}\right|^2 \delta(\omega-\epsilon_{r,k}) $. For simplicity, we assume the density of states $\rho_r$ and the tunneling amplitudes $t_r$ to be independent of energy, which implies constant tunneling rates $\Gamma_r$. Furthermore, we define $\Gamma=\Gamma_L+\Gamma_R$. The coupling of the leads via the reference arm is described by the dimensionless parameter $|t^\text{ref}|=2\pi|\tilde{t}|\sqrt{\rho_L\rho_R}$.

The quantum dot can be either empty or singly occupied with spin up or down.
Double occupancy is prohibited for large charging energy.
Therefore, the state of the system is described by a three-component vector of
the dot occupation probabilities ${\bf p}(N,t)=(p_0,p_\uparrow,p_\downarrow)^T$
under the condition that $N$ electrons have passed the system after time $t$.
(Spin symmetry, $p_\uparrow = p_\downarrow$, makes the problem effectively
two dimensional only.)
Its time evolution is governed by an $N$-resolved generalized master equation
\begin{equation}\label{eq:meq}
  \frac{d}{dt}{\bf p}(N,t)  =   \sum_{N'=-\infty}^{\infty} \int_0^{t} dt'
       {\bf W}(N-N',t-t') \cdot {\bf p}(N',t') .
\end{equation}
Transitions between the system states are described by the $3 \times 3$ matrix
${\bf W}(N-N',t-t')$.
In the Markovian limit, $\mathbf{W}(N-N',t-t') = \mathbf{W}(N-N')
 \delta(t-t')$, the probability vector at time $t_0$ becomes
${\bf p}(\chi,t_0) = \sum_N e^{i\chi N} {\bf p}(N,t_0) =
e^{\mathbf{W}(\chi) t_0} {\bf p}_0 $, where the initial state ${\bf p}_0$
does not depend on the counting field $\chi$.
Now we can perform a spectral decomposition of $\mathbf{W}(\chi) =
\sum_N e^{i\chi N}\mathbf{W}(N)$.
In the long-time limit the only contribution comes from the eigenvalue
$\lambda(\chi)$ of $\mathbf{W}(\chi)$ with the smallest absolute value of the real part.
Defining $P(N,t_0)=\sum_n p_n(N,t_0)$ we arrive at the result
\begin{equation}    \label{eq:Smarkov}
    S(\chi) = t_0 \, \lambda(\chi) .
\end{equation}

In general, non-Markovian corrections to this result--related to
a finite support of the kernels $\mathbf{W}(N-N',t-t')$ in time--may appear.
A generalization of Eq.~(\ref{eq:Smarkov}) that includes the non-Markovian
dynamics has been presented in Ref.~\onlinecite{braggio:2006}.
However, it has been shown there that non-Markovian corrections do not enter
the cumulant generating function for the lowest-order term of a perturbation
expansion in some small parameter.
For the quantum-dot ABI, the combination
$\alpha=|t^\text{ref}|\sqrt{\GL\GR}$ provides such a small parameter, i.e.,
the lowest-order contribution in $\alpha$ to the CGF does not contain
non-Markovian corrections.

To derive the kernels $\bf W$ of the $N$-resolved master equation, we make use of a diagrammatic real-time technique for the time evolution of the reduced density matrix formulated on a Keldysh contour.
For a detailed derivation of this diagrammatic language and the rules on how to calculate the value of a diagram we refer to Refs.~[\onlinecite{diagrams,technique,koeniggefen}].

The counting fields $\chi_r$ are incorporated in this approach by replacing the tunnel matrix elements in the Hamiltonian as $t_r \rightarrow t_r \, e^{\pm i\chi_r/2}$ and $\tilde t \rightarrow \tilde t \, e^{i\tilde \chi}$
with $\chi_L=-\chi_R=\tilde{\chi}/2=\chi/2$,
where the positive (negative) sign is taken for vertices on the upper (lower) branch of the Keldysh contour.

In the case of small dot-lead coupling, $\Gamma_r \ll k_BT$, all quantities may be expanded to first order in the tunnel couplings $\Gamma_{L,R}$. Interference effects are included in lowest order by continuing the expansion to the order $|t^\text{ref}|\Gamma_{L,R}$.
We expand the eigenvalue of $\bf W$ in the tunnel-coupling strengths and keep the three low-order
contributions of the cumulant generating function,
\begin{equation}
    S(\chi) =   S^{ (\Gamma) }(\chi) + S^{ (|t^\text{ref}|\Gamma) }(\chi,\phi)  +  S^{ (|t^\text{ref}|^2) } (\chi) \, .
\end{equation}
The first and the third part describe tunneling through the quantum dot and through the reference arm, respectively.
For the latter transport is known to be Poissonian in the tunneling limit. Interference is described by the second, flux-dependent term.

Starting from the Hamiltonian in Eq.~(\ref{eq:hamiltonian}) we calculate the kernel of the Master Equation~(\ref{eq:meq}) with the aforementioned diagrammatic real-time technique.
The kernel for a quantum dot without the reference arm $W^{(\Gamma)}$ has been previously obtained (see e.g.~Refs.~\onlinecite{bagrets:2003,braggio:2006}) and is repeated here, together with the
lowest-order interference term,
\begin{widetext}
\begin{eqnarray}
\label{eq:rates0}
  {\bf W}^{(\Gamma)} &=& \sum_{r=L,R} \Gamma_r
  \left(
  \begin{array}{ccc}
    -2 f_r(\epsilon) & [1-f_r(\epsilon)] e^{i\chi_r} &
    [1-f_r(\epsilon)] e^{i\chi_r}\\
    f_r(\epsilon) e^{-i\chi_r} & -1+f_r(\epsilon) & 0\\
    f_r(\epsilon) e^{-i\chi_r} & 0 & -1+f_r(\epsilon)
  \end{array}
  \right)
\\
\label{eq:rates1}
  {\bf W}^{(t^{\rm ref}\Gamma)} &=&
  |t^\text{ref}|\sqrt{\GL\GR} \cos{\phi}
  \left(
  \begin{array}{ccc}
    2 A(\chi) & 0 & 0 \\
    0 & A(\chi) & 0\\
    0 & 0 & A(\chi)
  \end{array}
  \right)
  +
  |t^\text{ref}|\sqrt{\GL\GR} \sin{\phi}
  \left(
  \begin{array}{ccc}
    - 2 B(\chi) & D(\chi) & D(\chi)\\
    C(\chi) & B(\chi) & 0\\
    C(\chi) & 0 & B(\chi)
  \end{array}
  \right)
\end{eqnarray}
where we used the abbreviations
\begin{eqnarray}
  \label{eq:A}
  A(\chi) &=& \int' \,\frac{\mathrm{d}\omega}{\pi} \,
  \frac{f_L(\omega)[1-f_R(\omega)](e^{i\chi}-1)
        +[1-f_L(\omega)]f_R(\omega)(e^{-i\chi}-1)}{\epsilon-\omega}
\\
  \label{eq:B}
  B(\chi) &=& \bLR\,e^{i\chi} - \bRL\,e^{-i\chi}
\\
  \label{eq:C}
  C(\chi) &=& f_L(\epsilon) [1-2f_R(\epsilon)] e^{i\frac{\chi}{2}} -  [1-2f_L(\epsilon)]f_R(\epsilon) e^{-i\frac{\chi}{2}}
\\
  \label{eq:D}
  D(\chi) &=& [1-2f_L(\epsilon)][1-f_R(\epsilon)] e^{i\frac{\chi}{2}} - [1-f_L(\epsilon)][1-2f_R(\epsilon)] e^{-i\frac{\chi}{2}}
  \, .
\end{eqnarray}
\end{widetext}
The processes appearing in the flux dependent part can be divided into two classes: Processes changing the dot state [$C(\chi)$ and $D(\chi)$] and processes that transfer charges without changing the dot state.
The latter may possess either off-resonant [$A(\chi)$] or resonant [$B(\chi)$] nature.
In contrast to the sequential-tunneling term, counting fields appear on the diagonal of the
kernel since there are processes that transfer charges through the entire interferometer leaving the
dot state unchanged.

\section{CGF for an interacting ABI} \label{sec:nonint}

The low-order contributions to the cumulant generating function
$S(\chi) =   S^{ (\Gamma) }(\chi) + S^{ (|t^\text{ref}|\Gamma) }(\chi,\phi) + S^{(|t^\text{ref}|^2)}(\chi)$
are found to be
\begin{eqnarray}
  S_{}^{ (\Gamma) }(\chi)
    &=& -t_0 \frac{\Gamma}{2} (F+1) \left( 1 - \sqrt{\mathcal{D}_{}} \right)                    \\
  \label{eq:Sinf1}
  \frac{ S_{}^{ (|t^\text{ref}|\Gamma) }(\chi,\phi) }{t_0 |t^\text{ref}|\sqrt{\GL\GR}}
    &=& \cos{\phi}\; A(\chi) \, \frac{2}{F+1}                                                       \nonumber\\
    &-& \frac{1}{2} \;  \cos{\phi}\; A(\chi) \, \Bigl[ 1-\frac{1}{\sqrt{\mathcal{D}_{}}} \Bigr] \frac{1-3F}{F+1} \nonumber\\
    &-& \frac{1}{2} \;  \sin{\phi}\; B(\chi) \, \Bigl[ 1-\frac{1}{\sqrt{\mathcal{D}_{}}} \Bigr] \\
  S_{}^{(|t^\text{ref}|^2)}(\chi)
    &=& t_0 \, eV \, |t^\text{ref}|^2 (e^{i\chi}-1)
\end{eqnarray}
where we have made use of the definitions
\begin{eqnarray}
  \mathcal{D}_{} &=& 1+8\frac{\GL\GR}{(\GL+\GR)^2} \, \frac{1}{(F+1)^2}
  \nonumber \\
  & & \times \left[ \bLRs(e^{i\chi}-1)
    \right. \nonumber \\ & & \quad \left.
    + \bRLs(e^{-i\chi}-1) \right]
  \\
  F &=& \frac{\Gamma_L f_L(\epsilon) + \Gamma_R f_R(\epsilon)}
  {\Gamma_L + \Gamma_R}
\end{eqnarray}
The second term, Eq.~(\ref{eq:Sinf1}), is the central result of this paper, while
the first\cite{bagrets:2003,braggio:2006} and third terms are well known.

For interacting electron systems, the notion of a transmission probability
$T(\omega)$ that contains all information about the full counting statistics
via the Levitov-Lesovik formula is, in general, not applicable anymore.
Incidentally such a notion still works for the lowest-order contribution $S_{}^{(\Gamma)}(\chi)$.
In fact, the non-interfering CGF for the interacting case can be reproduced by using that of
the noninteracting one but with rescaled coupling parameters.\cite{note_1}
For the interference part $S_{}^{(t^{\rm ref}\Gamma)}(\chi)$, however,
one could define a transmission probability $T(\omega)$ by writing the current
in the form $I=(e/h) \sum_\sigma \int d\omega T(\omega)
[f_L(\omega) - f_R(\omega)]$, but plugging this transmission probability
into the Levitov-Lesovik formula would not reproduce the higher cumulants.

Let us now discuss in detail how interaction changes the properties of
transport processes.
The interference part of the CGF contains terms for two kinds of transport
processes.
They differ both in their flux and voltage dependence.
The first type of processes is associated with interference between cotunneling
through the dot and direct tunneling from left to right.
These processes are of off-resonant nature and carry an even flux dependence.
They are described by the first two terms in
$S_{}^{(t^{\rm ref}\Gamma)}(\chi)$.
The first part looks similar to the CGF of the noninteracting problem.
The only change is the appearance of the factor $1/(F+1)$.
This prefactor describes the partial reduction of interference due to
spin-flip processes.
It has been predicted for the linear conductance,\cite{koeniggefen}
but it enters in the very same way for all cumulants and bias voltages.
In addition, however, there is a second part for these off-diagonal processes,
described by the second line of Eq.~(\ref{eq:Sinf1}), which is nonzero for
the second and higher cumulants.
Due to this term, the statistics is not bidirectional Poissonian anymore.
Furthermore, we see that the dependence on $F(\epsilon)$ becomes more
complicated.

The most significant effect of Coulomb interaction is, however, the appearance
of a second type of transport processes, described by the third line of Eq.~(\ref{eq:Sinf1}).
These processes are absent in the noninteracting case.
They are of on-resonant nature, and they carry an odd flux dependence.
As can be seen from the rates, Eq.~(\ref{eq:rates1}), some of them are related
to a change in the dot state, in contrast to the off-resonant cotunneling
terms that only occupy the dot virtually.
It should be noted that experimentally the $\sin{\phi}$- and $\cos{\phi}$-dependent parts can be easily distinguished from each other, either by performing a Fourier analysis for the cumulants, or by tuning the  flux such that only processes of one kind contribute to transport. With this in mind we proceed with studying separately the cumulant generating functions $S_\text{sin}$ and $S_\text{cos}$ (containing terms with $\sin{\phi}$- and $\cos{\phi}$-dependence respectively).

\section{Shot-Noise Regime} \label{sec:SN}

Let us now concentrate on the shot-noise regime, $eV=\mu_L-\mu_R\gg k_BT$.
In this case, $f_L(\epsilon)=1$ and $f_R(\epsilon)=0$, the transport voltage
dominates over thermal fluctuations and transport occurs mainly from left to right
lead both for resonant and non-resonant processes.
In particular, we are interested in the case of strongly asymmetric left and right tunnel couplings described by the ratio $ \gamma = \Gamma_L/\Gamma_R \ll 1 $.
As mentioned above, the flux-dependent contributions can be studied individually.
We therefore define the cumulants $\kappa^{(n)}_\text{cos,sin}$ as the parts of the cumulant with flux dependence $\cos{\phi},\sin{\phi}$.
Furthermore, we define generalized Fano factors as the quotients
$\kappa^{(n)}_\text{cos,sin}/I_\text{cos,sin}$ with
$I_\text{cos,sin} \equiv \kappa^{(1)}_\text{cos,sin}$ being the $\sin \phi$- and $\cos \phi$-dependent part
of the current, divided by $e$.

\begin{figure}\center
    \includegraphics[width=.95\columnwidth]{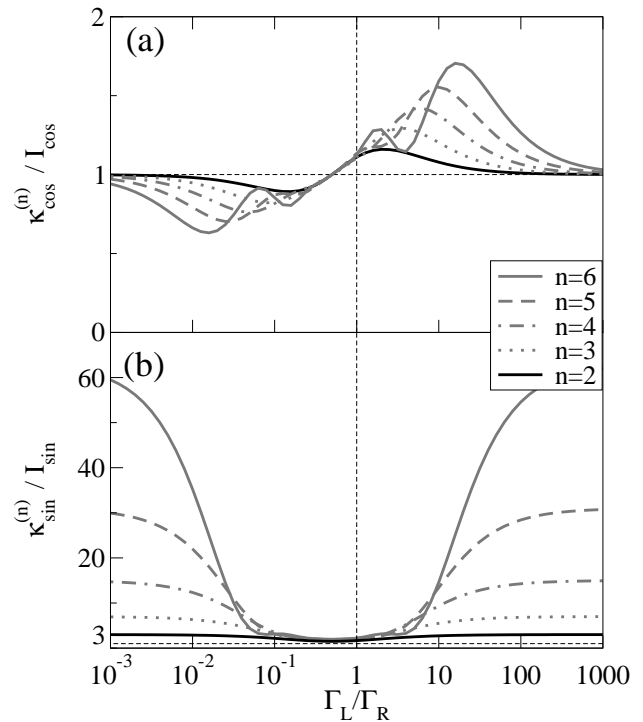}
    \caption{Generalized Fano factors $\kappa^{(n)}_\text{cos,sin}/I_\text{cos,sin}$ in the shot noise regime
    $\mu_L\gg\epsilon\gg\mu_R$. $\kappa^{(n)}_\text{cos,sin}$ is the $\cos{\phi}$/$\sin{\phi}$-dependent part of the $n$-th cumulant ($n=2$: noise, $n=3$: skewness, etc.).
	(a) Cosine-dependent generalized Fano factors, determined by off-resonant processes. The value $1$ is assumed
	at $\GL/\GR=1/2$.
	(b) Sine-dependent generalized Fano factors, determined by on-resonant processes.
	They approach $2^n-1$ for $\GL\gg\GR$ or $\GL\ll\GR$, while the minimum lies at $\GL/\GR=1/2$ and has the value $(2^n-1)/2^{n-1}$.
}
    \label{fig:Fbothn}
\end{figure}

Figure~\ref{fig:Fbothn}(a) shows the cosine-dependent generalized Fano factors as a function of
$\GL/\GR=\gamma$. In the extremely asymmetric case where one arm is almost pinched off
$\gamma\gg 1$ or $\gamma\ll 1$, the noise becomes Poissonian: $\kappa^{(n)}_\text{cos}=I_\text{cos}$.
This is not the case for the sine-dependent generalized Fano factors, Fig.~\ref{fig:Fbothn}(b).
They are enhanced and approach $ \kappa^{(n)}_\text{sin}/I^\text{sin}=2^n-1$ for $\gamma\gg 1$ or $\gamma\ll 1$.
For $n=2$, the Fano factor is 3.The transfer if $q$ charges in one or a sequence of multiple  elementary transport events is
associated\cite{fanogreaterone,andreevreflection,fanolessone} with a Fano factor
$ \kappa^{(2)} / \kappa^{(1)} = q $ and similar for higher cumulants ($ \kappa^{(3)}/ \kappa^{(1)} = q^2 $).\cite{levitov:2004}Within the framework of counting statistics, such transport processes with $q_i$ charges carry counting factors $e^{iq_i\chi}-1$.
The appearance of multiple counting factors with different $q_i$ implies that transport occurs through several elementary events with different charges.\cite{belzig:2005,cuevas:2003,johansson:2003}In our case, the Fano factor of $3$ does not hint at three charges per event, rather a combination of single and double charges are involved. Expanding the cumulant generating function in terms of $\gamma\ll 1$ reveals counting factors for single and double charge transfers:
\begin{eqnarray}\label{eq:CGFsn}
    S_{\infty} (\chi)
      &=& t_0 \,2 \Bigl[    \gamma\GR\,(e^{i\chi}-1)
          \nonumber \\
      & & +|t^\text{ref}|\GR\gamma^\frac{3}{2}\, \sin{\phi} \;
          e^{i\chi}(e^{i\chi}-1)   \label{eq:CGFsnSin}
          \nonumber \\
      & & +|t^\text{ref}|\GR\sqrt{\gamma}\cos{\phi} \;
          \frac{e^{i\chi}-1}{F+1}
          \frac{1}{\pi} \ln \left| \frac{\mu_L - \epsilon}{\mu_R - \epsilon} \right|
          \nonumber\\
      & & + eV |t^\text{ref}|^2 (e^{i\chi}-1)
          \Bigr]
 \, .
\end{eqnarray}
The first term describes transport trough the quantum dot in absence of the reference arm. Transport behavior is dominated by the smaller tunnel barrier $\GL=\gamma\GR$ and transport becomes Poissonian. The last line describes transport through the reference arm in absence of the dot and is also Poissonian.
The cosine part of the interference term (third term) is, for very asymetric tunnel couplings, Poissonian as well.
The sine part of the interference term, however, is different.
It contains a counting factor of $e^{i2\chi}$, which is responsible for the enhanced generalized Fano factors.
We note that this contribution is proportional to $\gamma^{3/2}$, i.e., one order higher in the asymmetry
$\gamma$ than the cosine term.

One may hope to identify the appearence of the $e^{i2\chi}$ term not only in the higher cumulants but also directly in the probability distribution $P(N,t_0)$ as an even-odd feature.
But once $t_0$ is large enough to get a reasonable number of transferred charges to identify the probability distribution, then the even-odd features from the sine part will be washed out by the other contributions
in Eq.~(\ref{eq:CGFsn}).

\section{QPC-Detector}  \label{sec:detector}

Current experimental techniques to measure the counting statistics of systems involving quantum dots employ quantum point contacts for detection of the dot's charge state.\cite{gustavsson:2006,fricke:2007,fujisawa:2006}
Depending on the charge state of the quantum dot, the nearby quantum point contact has a high or low transmission.
By a time-resolved measurement of the current through the quantum point contact one can monitor tunneling events that fill up or deplete the quantum dot.
For large source-drain voltages, electrons can only enter the dot from the source and leave to the drain electrode, which provides the information to translate the jumps in the QPC-current into the counting statistics of transport through the QD.

Such a direct correspondence is no longer available in multiply-connected geometries, such as the quantum-dot ABI considered in this paper, since there is more than one way to fill or empty the dot.
Therefore, it is an interesting question to ask whether and how much of the peculiarities of the FCS described in the previous section is accessible by measuring the charge of the quantum dot.
As the charge detection occurs with a finite bandwidth only, it was shown that the cumulants are systematically underestimated.\cite{naaman:2006,gustavsson:detector,flindt:2007}
In order to incorporate the detector in the description of the system we introduce the probability vector
${\bf p}=(p_{00},p_{\uparrow 0},p_{\downarrow 0},p_{01},p_{\uparrow 1},p_{\downarrow 1})$.\cite{naaman:2006,flindt:2007} The first index $n=0,\uparrow,\downarrow$ denotes the state of the dot and the second $m=0,1$ describes the state the detector believes the dot to be in. Upon change of the dot state the detector follows with a rate $\Gamma_D$. This is described by a master equation for the probabilities $p_{n,m}$:
\begin{widetext}
\begin{equation}\label{eq:ratesDetector}
    \frac{d}{dt}{\bf p}(t)  = \left(\begin{array}{cccccc}
        W_{0,0}	& W_{0,\up}			& W_{0,\dn}			& \Gamma_D			& 0			& 0         \\
        W_{\up,0}	& W_{\up,\up}-\Gamma_D	& 0					& 0					& 0			& 0         \\
        W_{\dn,0}	& 0					& W_{\dn,\dn}-\Gamma_D	& 0					& 0			& 0         \\
        0		& 0					& 0					& W_{0,0}-\Gamma_D	& W_{0,\up}	& W_{0,\dn}   \\
        0		& \Gamma_D e^{i\chi}	& 0					& W_{\up,0}			& W_{\up,\up}	& 0         \\
        0		& 0					& \Gamma_D e^{i\chi}	& W_{\dn,0}			& 0			& W_{\dn,\dn}   \\
                            \end{array}\right) {\bf p}(t) .
\end{equation}
\end{widetext}
The rates $W_{i,j}$ are the rates of the system in absence of the detector. For our system they are given by Eqns.~(\ref{eq:rates0}) and (\ref{eq:rates1}) taken in the shot noise regime for vanishing counting field $ W_{i,j} = \left[ W_{i,j}^{(\Gamma)}+W_{i,j}^{(|t^\text{ref}|\Gamma)} \right]_{{eV \gg kT}\atop{\chi=0}} $.
We introduced the counting factor $e^{i\chi}$ for the transition from $p_{1,0}$ to $p_{1,1}$, i.e.~when the charge on the dot is detected.

The counting statistics for the detector can be obtained in the same way as before by taking the eigenvalue with the lowest negative real part. The lowest-order generating function for transport through the quantum dot has been calculated before.\cite{flindt:2007} There it was also discussed that in the limit of infinite bandwidth $\Gamma_D\rightarrow\infty$ the generating function for a quantum dot\cite{bagrets:2003} is recovered.

One may be worried that interference is destroyed by detecting the electrons on the dot.
This is, however, only true for open quantum-dot ABI for which measuring the dot charge provides a which-path information.\cite{buks:1998}
In closed interferometers a measurement of the dot charge does not yield path information, because paths encircling the flux several times are possible.\cite{khym:2006,chang:2007}
Furthermore, even without allowing for such higher winding numbers the knowledge of the electrons being on the quantum dot does not include the knowledge of the path along which it leaves: The electron might tunnel directly to the drain lead or first go back to the source virtually and then tunnel to the drain via the reference arm. These processes are exactly those described by the resonant terms $C(\chi)$ and $D(\chi)$.

As a consequence we find a flux-dependent correction to transport through the dot. For a finite bandwidth the generating function has a complicated dependence on $\Gamma_D$ which we therefore do not show. It describes a reduction of all moments due to the finite bandwidth. However, for infinite bandwidth $\Gamma_D\rightarrow\infty$ this generating function simplifies to
\begin{widetext}
\begin{equation}\label{eq:Sdetector}
  \left.S_{\text{QPC},U=\infty}^{(|t^\text{ref}|\Gamma)}\right|_{\Gamma_D\rightarrow\infty}   =   -|t^\text{ref}|\sqrt{\GL\GR}\sin{\phi} \; \left( 2\,\frac{\GL-\GR}{2\GL+\GR} \frac{e^{i\chi}-1}{\sqrt{\Dinf}}      + \frac{1}{2}\left(1-\frac{1}{\sqrt{\Dinf}}\right) \right)
\end{equation}
\end{widetext}

This result has the following properties.
First, no cosine terms appear. This is clear, because the detector is insensitive to the off-resonant cotunneling processes $A(\chi)$ which go along with only a virtual occupation of the dot. In addition, the detector is insensitive to the {\it resonant} contribution $B(\chi)$ to the cotunneling processes, as they also preserve the dot state.
Correspondingly, the same statistic can be obtained if we replace $A(\chi)$ by $A(0)$ and $B(\chi)$ by $B(0)$ in Eq.~(\ref{eq:rates1}) instead of solving the detector model Eq.~(\ref{eq:ratesDetector}).
In this case it becomes apparent that the flux-dependent contribution may be understood as a correction to the tunneling rates $\Gamma_{L,R}$: The rates in the kernel Eqns.~(\ref{eq:rates0},\ref{eq:rates1}) become
    $
        W_{1,0} =   \Gamma_L + |t^\text{ref}|\sqrt{\Gamma_L\Gamma_R} \sin{\phi}
    $
    and
    $
        W_{0,1} =   \Gamma_R - |t^\text{ref}|\sqrt{\Gamma_L\Gamma_R} \sin{\phi}
    $.
Expanding the non-interfering generating function for a QD with these rescaled rates in terms of $|t^\text{ref}|$ yields the first part of Eq.~(\ref{eq:Sdetector}).
As it is solely caused by rescaled coupling parameters a similar term is also present in the statistics of a noninteracting system. There however it constitutes the entire generating function.
The presence of Coulomb interaction causes the second term of Eq.~(\ref{eq:Sdetector}).

\begin{figure}\center
    \includegraphics[width=.95\columnwidth]{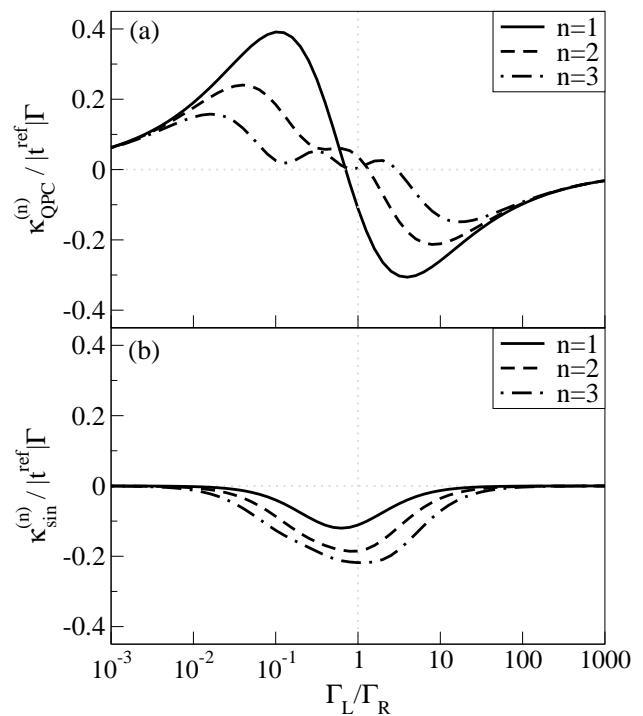}
    \caption{Logarithmic plots of the $n$-th cumulants $\kappa^{(n)}_\text{QPC}$ (in presence of the QPC-detector, black, from Eq.~(\ref{eq:Sdetector})) and $\kappa^{(n)}_\text{sin}$ (without the QPD-detector, grey, from Eq.~(\ref{eq:Sinf1})) in the shot noise regime $\mu_L\gg\epsilon\gg\mu_R$ (such that $f_L=1,f_R=0$). The value $1$ is plotted for reference.
}
    \label{fig:QPCmoments}
\end{figure}

In Fig.~(\ref{fig:QPCmoments}) we plot the flux-dependent corrections to the cumulants.
These corrections change sign for specific values of $\Gamma_L/\Gamma_R$.

In conclusion, we find that measuring the quantum-dot charge by a nearby QPC does, indeed, provide some information of the FCS of the transport through the quantum-dot ABI.
Interference processes that do not change the occupation of the dot remain undetected.
This includes the off-resonant interference contributions with an even flux dependence and part of the
on-resonant part with an odd flux dependence.
The on-resonant interference processes that are accompanied with a change of the dot state, though,
could be detected in that way.

\section{Conclusion}    \label{sec:conclusion}

We examined the counting statistics of a quantum-dot ABI in the limit of weak tunnel couplings.
The cumulant generating function allows to clearly identify two types of transport processes with different flux dependences:
First, there are off-resonant processes with an even flux dependence.
They appear in the absence of Coulomb interaction as well.
But second (for finite bias voltages) on-resonant processes with an odd dependence on flux contribute to transport, which are not present for vanishing Coulomb interaction. Due to the different dependence on flux both kinds of processes can be studied separately.
We found that the on-resonant $\sin{\phi}$-dependent term obeys an interesting statistics.
This is most dramatically seen in the limit of very asymmetric tunnel couplings of the dot to the leads.
While the off-resonant processes obey Poissonian statistics in this limit, the resonant processes show strongly super-Poissonian behavior.
Finally, we proved that the existence of some of these on-resonant $\sin \phi$ dependent interference contributions could be proven by measuring the quantum-dot charge by a nearby quantum point contact.

\acknowledgments

We acknowledge useful discussions with W.~Belzig and financial support from DFG via
GRK 72, MIUR-PRIN and EU-RTNNANO.

\appendix


\begin{thebibliography}{99}

    \bibitem{levitov:1996}
        L.~S.~Levitov and G.~B.~Lesovik,
        Pis'ma Zh.~E´ksp.~Teor.~Fiz.~{\bf 58}, 225 (1993) [JETP Lett.~{\bf 58}, 230 (1993)];
        L.~S.~Levitov, H.-W.~Lee, and G.~B.~Lesovik,
        J.~Math.~Phys.~{\bf 37}, 4845 (1996).
    \bibitem{QNmesPhys}
        A good overview is found in {\it Quantum Noise in Mesoscopic Physics},
        ed.~by~Yu.~V.~Nazarov, NATO Science Series, Kluwer (2003).
    \bibitem{nazarov:2002}
        Yu.~V.~Nazarov and D.~A.~Bagrets,
`        Phys.~Rev.~Lett.~{\bf 88}, 196801 (2002).
    \bibitem{boltzmannlangevin}
        K.~E.~Nagaev,
        Phys.~Rev.~B {\bf 66}, 075334 (2002).
    \bibitem{KeldyshGF}
        W.~Belzig and Yu.~V.~Nazarov,
        Phys.~Rev.~Lett.~{\bf 87}, 067006 (2001);
        Yu.~V.~Nazarov and M.~Kindermann,
        Eur.~Phys.~J.~B~{\bf 35}, 413 (2003).
    \bibitem{stochasticPathint}
        S.~Pilgram, A.~N.~Jordan, E.~V.~Sukhorukov, and M.~Buttiker,
        Phys.~Rev.~Lett.~{\bf 90}, 206801 (2003);
        A.~N.~Jordan, E.~V.~Sukhorukov, and S.~Pilgram,
        J.~Math.~Phys.~{\bf 45}, 4386 (2004).
    \bibitem{braggio:2006}
        A.~Braggio, J.~K\"onig and R.~Fazio,
        Phys.~Rev.~Lett.~{\bf 96}, 026805 (2006).
    \bibitem{flindt:2008}
        C.~Flindt, T.~Novotn\'y, A.~Braggio, M.~Sassetti, and A.-P.~Jauho,
        Phys.~Rev.~Lett.~{\bf 100}, 150601 (2008).
    \bibitem{emary:2007}
        C.~Emary, D.~Marcos, R.~Aguado, and T.~Brandes,
        Phys.~Rev.~B {\bf 76}, 161404(R) (2007).
    \bibitem{TunnelDots}
        D.~A.~Bagrets, Y.~Utsumi, D.~S.~Golubev, and G.~Schoen,
        Fortschritte der Physik {\bf 54}, 917 (2006).
    \bibitem{bagrets:2003}
        D.~A.~Bagrets and Y.~V.~Nazarov,
        Phys.~Rev.~B {\bf 67}, 085316 (2003).
    \bibitem{kiesslich:2006}
        G.~Kie\ss lich, P.~Samuelsson, A.~Wacker and E.~Sch\"oll,
        Phys.~Rev.~B {\bf 73}, 033312 (2006).
    \bibitem{gogolin:2006}
        A.~O.~Gogolin and A.~Komnik,
        Phys.~Rev.~Lett.~{\bf 97}, 016602 (2006).
    \bibitem{schmidt:2007}
        T.~L.~Schmidt, A.~O.~Gogolin, and A.~Komnik,
        Phys.~Rev.~B~{\bf 75}, 235105 (2007).    \bibitem{SCleads}
        W.~Belzig,
        in: {\it Quantum Noise in Mesoscopic Physics},
        ed.~by~Yu.~V.~Nazarov, NATO Science Series, Kluwer (2003).    \bibitem{heikkila}
        M.~A.~Laakso, P.~Virtanen, F.~Giazotto, and T.T.~Heikkil\"a,
        Phys.~Rev.~B~{\bf 75},094507 (2007).
    \bibitem{choi:2001}
        M.-S.~Choi, F.~Plastina and R.~Fazio,
        Phys.~Rev.~Lett.~{\bf 87}, 116601 (2001);
        M.-S.~Choi, F.~Plastina and R.~Fazio,
        Phys.~Rev.~B~{\bf 67}, 045105 (2003).
    \bibitem{NEMS}
        C.~Flindt, T.~Novotny, and A.-P.~Jauho,
        Europhys.~Lett.~{\bf 69} 475 (2005);
        F.~Pistolesi,
        Phys.~Rev.~B~{\bf 69}, 245409 (2004).

    \bibitem{fanogreaterone}
        Ya.~M.~Blanter, "Recent Advances in Studies of Current Noise",
        to be published in: {\it Springer Lecture Notes}, edited by Ch.~R{\"o}thig, G.~Sch{\"o}n and M.~Vojta,
        preprint: arXiv:cond-mat/0511478v2.
    \bibitem{andreevreflection}
        F.~Lefloch, C.~Hoffmann, M.~Sanquer, and D.~Quirion,
        Phys.~Rev.~Lett.~{\bf 90}, 067002 (2003);        D.~Averin and H.~T.~Imam,
        Phys.~Rev.~Lett.~{\bf 76}, 3814 (1996);
        P.~Dieleman, H.~G.~Bukkems, T.~M.~Klapwijk, M.~Schicke, and K.~H.~Gundlach,
        Phys.~Rev.~Lett.~{\bf 79}, 3486 (1997).

    \bibitem{fanolessone}
        R.~de-Picciotto, M.~Reznikov, M.~Heiblum, V.~Umansky, G.~Bunin and D.~Mahalu
        Nature {\bf 389}, 162 (1997);
        L.~Saminadayar and D.~C.~Glattli, Y.~Jin and B.~Etienne,
        Phys.~Rev.~Lett.~{\bf 79}, 2526 (1997).

    \bibitem{cuevas:2003}
        J.~C.~Cuevas and W.~Belzig,
        Phys.~Rev.~Lett.~{\bf 91}, 187001 (2003),
        Phys.~Rev.~B {\bf 70}, 214512 (2004).
    \bibitem{johansson:2003}
        G.~Johansson, P.~Samuelsson, A.~Ingerman,
        Phys.~Rev.~Lett.~{\bf 91}, 187002 (2003).
    \bibitem{belzig:2005}
        W.~Belzig,
        Phys.~Rev.~B {\bf 71}, 161301(R) (2005).
    \bibitem{vanevic:2007}
        M.~Vanevic, Y.~V.~Nazarov, W.~Belzig,
        Phys.~Rev.~Lett.~{\bf 99}, 076601 (2007).

    \bibitem{hist3rdmoment}     
        B.~Reulet, J.~Senzier, and D.~E.~Prober,
        Phys.~Rev.~Lett.~{\bf 91}, 196601 (2003);
        Yu.~Bomze, G.~Gershon, D.~Shovkun, L.~S.~Levitov and M.~Reznikov,
        Phys.~Rev.~Lett.~{\bf 95}, 176601 (2005).

    \bibitem{qubitdetector}
        R.~J.~Schoelkopf, A.~A.~Clerk, S.~M.~Girvin, K.~W.~Lehnert, and M.~H.~Devoret,
        in {\it Quantum Noise in Mesoscopic Physics}, edited by Yu.~V.~Nazarov (Kluwer, Dordrecht, 2003).

    \bibitem{thermalescape}
        B.~Huard, H.~Pothier, N.O.~Birge, D.~Esteve, X.~Waintal, and J.~Ankerhold
        Ann.~Phys.~{\bf 16}, 736 (2007).
    \bibitem{cooperpairtunneling}   
        R.~K.~Lindell, J.~Delahaye, M.~A.~Sillanp\"a\"a, T.~T.~Heikkil\"a, E.~B.~Sonin, and P.~J.~Hakonen,
        Phys.~Rev.~Lett.~93, 197002 (2004);
        T.~T.~Heikkil\"a, P.~Virtanen, G.~Johansson, and F.~K.~Wilhelm,
        Phys.~Rev.~Lett.~93, 247005 (2004);
        A.V.~Timofeev, M.~Meschke, J.~T.~Peltonen, T.~T.~Heikkil\"a, and J.~P.~Pekola,
        Phys.~Rev.~Lett.~{\bf 98}, 207001 (2007).

    \bibitem{MQTfourthcumulant}     
        J.~Ankerhold and H.~Grabert,
        Phys.~Rev.~Lett.~{\bf 95}, 186601 (2005).

    \bibitem{thresholddetectors}
        J.~Tobiska and Yu.~V.~Nazarov,
        Phys.~Rev.~Lett.~{\bf 93}, 106801 (2004);
        J.~P.~Pekola,
        Phys.~Rev.~Lett.~{\bf 93}, 206601 (2004)

    \bibitem{gustavsson:2006}  
        S.~Gustavsson, R.~Leturcq, B.~Simovic, R.~Schleser, P.~Studerus, T.~Ihn, K.~Ensslin, D.~C.~Driscoll, and A.~C.~Gossard,
        Phys.~Rev.~B {\bf 74}, 195305 (2006).
    \bibitem{fricke:2007}       
        C.~Fricke, F.~Hohls, W.~Wegscheider, and R.~J.~Haug,
        Phys.~Rev.~B~{\bf 76}, 155307 (2007).

    \bibitem{fujisawa:2006}
        T.~Fujisawa, T.~Hayashi, R.~Tomita, Y.~Hirayama,
        Science~{\bf 312}, 1634 (2006).

    \bibitem{forster}
        S.~Pilgram, P.~Samuelsson, H.~F\"orster, M.~B\"uttiker,
        Phys.~Rev.~Lett.~{\bf 97}, 066801 (2006);
        H.~F\"orster, P.~Samuelsson, S.~Pilgram, M.~B\"uttiker,
        Phys.~Rev.~B {\bf 75}, 035340 (2007).
    \bibitem{buks:1998}   
        E.~Buks, R.~Schuster, M.~Heiblum, D.~Mahalu and V.~Umansky,
        Nature~{\bf 391}, 871 (1998).
    \bibitem{khym:2006}
        G.~Luck~Khym and K.~Kang,
        Phys.~Rev.~B~{\bf 74}, 153309 (2006).
    \bibitem{ChargeDetector}    
        V.~Moldoveanu, M.~Tolea, and B.~Tanatar,
        Phys.~Rev.~B~{\bf 75}, 045309 (2007).
    \bibitem{EdgeStates}
        D.~Rohrlich, O.~Zarchin, M.~Heiblum, D.~Mahalu, and V.~Umansky,
        Phys.~Rev.~Lett.~{\bf 98}, 096803 (2007).
    \bibitem{PhaseDetector}
        D.~Sprinzak, E.~Buks, M.~Heiblum, and H.~Shtrikman,
        Phys.~Rev.~Lett.~{\bf 84}, 5820 (2000).
    \bibitem{KondoDephasing}
        D.~Sprinzak, Y.~Ji, M.~Heiblum, D.~Mahalu, and H.~Shtrikman,
        Phys.~Rev.~Lett.~{\bf 88}, 176805 (2002);
        M.~Avinun-Kalish, M.~Heiblum, A.~Silva, D.~Mahalu, and V.~Umansky,
        Phys.~Rev.~Lett.~{\bf 92}, 156801 (2004).

    \bibitem{PhaseLapseOriginal}
        A.~Yacoby, M.~Heiblum, D.~Mahalu, and H.~Shtrikman,
        Phys.~Rev.~Lett.~{\bf 74}, 4047 (1995).
        R.~Schuster, E.~Buks, M.~Heiblum, D.~Mahalu, V.~Umansky ,and H.~.Shtrikman
        Nature~{\bf 385}, 417 (1997).
    \bibitem{PhaseLapseNewExps1}
        M.~Avinun-Kalish, M.~Heiblum, O.~Zarchin, D.~Mahalu and V.~Umansky
        Nature~{\bf 436}, 529 (2005).
    \bibitem{PhaseLapseNewExps2}
        A.~Aharony, O.~Entin-Wohlman, T.~Otsuka, S.~Katsumoto, H.~Aikawa, and K.~Kobayashi,
        Phys.~Rev.~B~{\bf 73}, 195329 (2006);
    \bibitem{kobayashi:2004}
        H.~Aikawa, K.~Kobayashi, A.~Sano, S.~Katsumoto, and Y.~Iye,
        Phys.~Rev.~Lett.~{\bf 92}, 176802 (2004).
    \bibitem{unexpectedperiodicity}
        A. Yacoby, R. Schuster, and M. Heiblum,
        Phys.~Rev.~B~{\bf 53}, 9583(1996).

    \bibitem{PhaseLapseAnalysis}
        A.~L.~Yeyati and M.~B\"uttiker,
        Phys.~Rev.~B~{\bf 52}, R14360 (1995);
        G.~Hackenbroich and H.~A.~Weidenm\"uller,
        Phys.~Rev.~Lett.~{\bf 76}, 110 (1996);
        G.~Hackenbroich and H.~A.~Weidenm\"uller,
        Phys.~Rev.~B~{\bf 53}, 16379 (1996);
        Y.~Oreg and Y.~Gefen,
        Phys.~Rev.~B~{\bf 55}, 13726 (1997);
        H.~A.~Weidenm\"uller,
        Phys.~Rev.~B~{\bf 65}, 245322 (2002);
        O.~Entin-Wohlman, A.~Aharony, Y.~Imry and Y.~Levinson
        J.~Low~Temp.~Phys.~{\bf 126}, 1251 (2002);
        A review of theoretical work can be found in: Y.~Gefen,
        in "Quantum Interferometry with Electrons: Outstanding Challenges", edited by I.V.~Lerner, B.~L.~Altshuler, V.~I.~Fal'ko, and T.~Giamarchi
        (Kluwer, Dordrecht, 2002), p.~13.
    \bibitem{hofstetter}
        W.~Hofstetter, J.~K\"onig, and H~Schoeller,
        Phys.~Rev.~Lett.~{\bf 87}, 156803 (2001).
    \bibitem{bruder:1996}
        C.~Bruder, R.~Fazio, and H.~Schoeller,
        Phys.~Rev.~Lett.~{\bf 76}, 114 (1996).
    \bibitem{pala:2004}
        M.~G.~Pala and G.~Iannaccone,
        Phys.~Ref.~Lett.~{\bf 93}, 256803 (2004).

    \bibitem{kubala}
        B.~Kubala and J.~K\"onig,
        Phys.~Rev.~B~{\bf 67}, 205303 (2003).

    \bibitem{onsager}
        L.~Onsager,
        Phys.~Rev.~{\bf 38}, 2265 (1931);
        H.~B.~G.~Casimir,
        Rev.~Mod.~Phys.~{\bf 17}, 343 (1945).

    \bibitem{koeniggefen}
        J.~K\"onig, and Y.~Gefen,
        Phys.~Rev.~Lett.~{\bf 86}, 3855 (2001);
        Phys.~Rev.~B~{\bf 65}, 045316 (2002).

    \bibitem{diagrams}
        J.~K\"onig, H.~Schoeller, and G.~Sch\"on,
        Phys.~Rev.~Lett.~{\bf 76}, 1715 (1996);
        J.~K\"onig, J.~Schmid, H.~Schoeller, and G.~Sch\"on,
        Phys.~Rev.~B {\bf 54} 16820 (1996).
    \bibitem{technique}
        H.~Schoeller, in {\it Mesoscopic Electron Transport}, edited by L.~L.~Sohn, L.~P.~Kouwenhoven, and G.~Sch\"on (Kluwer, Dordrecht, 1997);
        J.~K\"onig, {\it Quantum Fluctuations in the Single-Electron Transistor} (Shaker, Aachen, 1999).

    \bibitem{note_1}
    This rescaling is given by
    $  \Gamma_{L,R}^{U=0} \rightarrow \frac{\Gamma(F+1)}{2}\left( 1 \pm \sqrt{ 1-\frac{8\GL\GR}{\Gamma^2(F+1)^2} } \right)  $.

    \bibitem{levitov:2004}
        L.~S.~Levitov and M.~Reznikov,
        Phys.~Rev.~B~{\bf 70}, 115305 (2004).


    \bibitem{gustavsson:detector}
        S.~Gustavsson, R.~Leturcq, T.~Ihn, K.~Ensslin, M.~Reinwald, and W.~Wegscheider,
        Phys.~Rev.~B~{\bf 75}, 075314 (2007).

    \bibitem{naaman:2006}
        O.~Naaman and J.~Aumentado,
        Phys.~Rev.~Lett.~{\bf 96}, 100201 (2006).

    \bibitem{flindt:2007}
        C.~Flindt, A.~Braggio, and T.~Novotn{\'y},
        in {\it Noise and Fluctuations: 19th International Conference on Noise and Fluctuations.}
        AIP Conf.~Proc.~No.~922 (AIP, New York, 2007), p.~531.

    \bibitem{gustavsson:nanowire}
        S.~Gustavsson, I.~Shorubalko, R.~Leturcq, S.~Sch\"on, and K.~Ensslin,
        Appl.~Phys.~Lett.~{\bf 92}, 152101 (2008).

    \bibitem{chang:2007}
        D.-I.~Chang, G.~L.~Khym, K.~Kang, Y.~Chung, H.-J.~Lee, M.~Seo, M.~Heiblum, D.~Mahalu, and V.~Umansky,
        Nat.~Phys., {\bf 4}, 205 (2008).

\end{thebibliography}
\end{document}